\documentclass[runningheads]{llncs}
\usepackage[T1]{fontenc}
\usepackage{orcidlink}
\usepackage{graphicx}
\begin{document}
\title{Santa Clara 3D: Digital Reconstruction and Storytelling of a Francoist Concentration Camp}

\author{Stinne Zacho\and
Chris Hall\and
Jakob Kusnick\orcidlink{0000-0002-1653-6614}\and
Stefan Jänicke\orcidlink{0000-0001-9353-5212}
}

%
\authorrunning{Zacho et al.}
%
\institute{Department of Mathematics and Computer Science
\\University of Southern Denmark, \\
Campusvej 55, 5230, Odense, Denmark\\
}
%
%
%
%
%
\maketitle              
\begin{abstract}
This paper explores the potential of digital reconstruction and interactive storytelling to preserve historically suppressed sites. The main objective of an interdisciplinary team of data scientists from the MEMORISE project and associates of the memory association Asociación Recuerdo y Dignidad was to preserve the memory of the Francoist Santa Clara concentration camp in Soria, Spain, through the use of digital technology. 
Combining archival research, 3D modelling, 360° photography, and web development, a prototype digital platform was created to visualise the transformation of the site across three historical phases: its origin as a convent, its use as a Francoist concentration camp, and its present-day condition. The platform allows users to navigate through spatial and temporal layers. Clickable media markers encourage exploration and interaction. Drawing on principles of participatory design, narrative visualisation, and open-ended user engagement, the project demonstrates how digital tools can support memory work, public engagement, and historical reflection. Our low-cost concept is especially adaptable to other physical sites that have been erased or forgotten.
\end{abstract}
\section{Introduction}
The use of digital tools is introducing new ways to preserve and share cultural memory,  especially in places where the physical signs of the past have been removed or have changed over time. In the growing field of digital heritage, 3D reconstructions and web-based immersive experiences are now being used to engage the public, support learning, and keep memories alive~\cite{gomes20143d,rodriguez2024systematic}. These approaches are useful for remembering places associated with violence or repression, where traditional memorials may be absent and narratives remain suppressed.
Developed in collaboration with a local memory association Asociación Recuerdo y Dignidad and as part of the MEMORISE project~\cite{janicke2024memorise}, we explore how digital reconstruction can make hidden or neglected histories more accessible and visible to a broader audience. MEMORISE uses a wide range of digital tools, including interactive storytelling, 3D reconstructions and participatory design, to enhance memory work related to Nazi persecutions.

A main goal of this work is to create a low-cost pipeline for reconstructing former landmarks of persecution, to create at least a virtual space to tell the stories of their victims. Our focus is the Santa Clara Park in Soria, Spain, which functioned as a concentration camp during the early phase of the Spanish Civil War. Today, very few visible traces remain of the camp’s structures, and its role in the history of Francoist repression is not widely recognized. We combine historical research with interactive digital design to create a web-based platform, featuring digital 3D reconstructions, 360° photographs and historical context. This allows users to explore the site’s transformation across three historical phases: its origin as a convent, its use as a concentration camp and its present-day condition. Our main contributions are:
\begin{itemize}
 \item A web-based platform for exploring historical information about Santa Clara, from its founding to the present day.
 \item A virtual 3D reconstruction of the concentration camp at Santa Clara Park, based on historical sources and spatial interpretation.
 \item A reflection on how digital reconstructions can contribute to memory work in places without physical remains.
 \item A resource for raising awareness, educating, and informing the public about the historical significance of the site.
 \item A low-cost reproducible pipeline adaptable to relatable historical sites.
\end{itemize} 

We adopt a hybrid storytelling approach grounded in the research on author-driven and reader-driven data storytelling~\cite{segel2010narrative}. Allowing users to follow a structured timeline but also to explore freely based on their interest offers both educational value and emotional impact. To ensure that the final product is meaningful to both experts and the public, we followed a participatory visualisation design approach, which seeks to connect research with public engagement by actively involving users and stakeholders in the design process~\cite{janicke2020participatory}.   

Through this work, we contribute to the field of digital heritage by showing how interactive virtual environments can aid in preserving and conveying the stories of neglected historical sites. By combining digital reconstruction with historical context and collaborating with local memory organisations, this thesis seeks to offer innovative approaches to preserving, sharing, and understanding sites of repression.
We argue that digital reconstructions, when paired with collaborative storytelling methods, can effectively make suppressed histories more visible, engaging, and educational for diverse audiences.

\section{Related Work}

Digital technology can support a better understanding and remembrance of traumatic historical events~\cite{suhardjono2023use} such as the Holocaust or political repression during the Spanish Civil War. The digital reconstruction and storytelling of the Santa Clara Concentration Camp intersects with three main areas of related research. First, related projects concerning digital museums and virtual tours and how they use these platforms to communicate. Secondly, the use of digital reconstruction of historical sites and how they are valuable for reviving lost or transformed historical sites. Finally, the importance of digital storytelling as an asset for educating and conveying historical topics. These areas of research are reflected in the MEMORISE project~\cite{janicke2024memorise}, which combines 3D visualisation and interactive storytelling techniques to support Holocaust memory--approaches that were also used to develop Santa Clara 3D.

\subsection{Digital Museums and Virtual Tours}

In recent years, digital museums and virtual tours have become a valuable tool for educating and engaging users~\cite{antonaci2013virtual,chiao2018examining}. In particular, Covid-19 raised awareness on the need to develop solutions that enable remote access to cultural heritage through virtual museums~\cite{pourmoradian2021museum,li2022evaluation,meinecke2022towards}. 
The creation of guided tours in digital twins of existing memorial sites related to persecution has led to easier access to historical sites, and provided a chance to reach wider audiences. One example is the digital twin of Block 15 of the Haidari concentration camp~\cite{benardou2022first}, a building that still exists without being accessible to the public due to its location on a military base. Another example is the Yad Vashem 360° virtual tour~\cite{yadvashem}, which offers an immersive experience of the museum and guides the public through the history of Holocaust. By adding annotations to the 360° images, the viewer is presented with historical videos, survivor testimonies and animated models all contributing to an engaging interactive experience, which can support critical thinking and emotional engagement.
Similarly, the Terezín Memorial~\cite{terezin} and Mauthausen~\cite{mauthausen} virtual tours are examples of digital environments that use annotations to provide the user with information while preserving the memory of these historically significant places.
These examples inspire the design of the Santa Clara platform, particularly in using 360° visuals and annotations to create an accessible and emotionally engaging digital environment.

\subsection{Digital Reconstruction of Historical Sites}

The digital reconstruction of historical sites has been proven to be a very valuable tool for reviving sites that have otherwise been lost or transformed over time. These reconstructions introduce the possibility to visualise buildings, landscapes and architectural details that are no longer visible today~\cite{liritzis20213d}.
An example of this is the Virtual Reconstruction of Bergen-Belsen Concentration Camp~\cite{oliva2015recovering}. The model was developed using historical photographs, maps and survivor testimonies. This 3D reconstruction gives visitors the opportunity to learn about its past layout, view areas within the reconstruction and gain insight into the camp’s history. Similar projects include the Natzweiler Struthof concentration camp project~\cite{koehl2021memory}, which used methods like laser scanning, photogrammetry and UAV imagery and the 3D digital reconstruction of Auschwitz-Birkenau~\cite{jaskot2017architecture} that was developed using SketchUp.

The potential of using digital tools such as virtual reality and other technologies for enhancing Holocaust memory and education are emphasized in the article "Virtualising Memoryscapes: Guidelines for the Digital Holocaust Memory Project"~\cite{reframe}. The article argues how the integration of geolocated historical data and storytelling can create immersive educational experiences, a better understanding and deeper engagement.
Although Santa Clara is not a Holocaust site, the digital reconstruction draws on similar methods, combining archival materials, historical maps, and digital modelling, to recover a site where physical traces have mostly disappeared.

\subsection{Digital Storytelling}

Digital storytelling methods have proven to be a valuable asset for engaging audiences with complex historical subjects like the Holocaust. The combination of narrative techniques combined with interactive technologies enhances empathy, understanding and remembrance of these difficult topics. Dimensions in Testimony~\cite{dimensions} for example uses natural language processing to interact with pre-recorded interviews of Holocaust survivors. The user can engage in seemingly real-time conversations with the victims, making this an innovative way of preserving and presenting survivors' testimonies. Similarly, the Anne Frank VR Tour~\cite{annefrank} shows how interactive storytelling can create a powerful learning tool when combined with personal testimonies and reconstructed environments. 

The Spanish Civil War Memory Project~\cite{sandiego} is another example of an interactive platform for exploring audiovisual testimonies from survivors of the Francoist regime. By telling their stories, the project aims to create attention about the victims of the Francoist repression, as much of the history and documentation from that time has been destroyed or forgotten.

In a survey conducted by Meffert et al.~\cite{meffert2024survey} different ways of using digital storytelling for presenting heritage related to Nazi persecutions are explored. Methods such as linear and non-linear storytelling, interactive timelines and spatial navigation are highlighted as effective ways to make difficult historical topics more engaging and easier to understand. These insights are very relevant to the design of this project, as it emphasizes the importance of interactivity, spatial exploration and contextual information.

Furthermore, the structure of this platform also takes inspiration from the concepts of author-driven and reader-driven storytelling~\cite{segel2010narrative}. These concepts offer a balance between narrative control and user flexibility and the article emphasises the benefits of combining these methods to enhance clarity and engagement. In this project, a hybrid storytelling approach is used, allowing the viewer to follow the timeline slider, or explore the 3D models, geo-spatial maps and narratives non-linearly, allowing both guided learning and open-ended interactions with the history, highlighting the involved multimodal entities such as persons, objects and various sets of them with their inter-connections~\cite{kusnick2024every}.

These initiatives reflect the growing importance of digital storytelling in both holocaust education and the broader memorialisation of political violence. The combination of technology with spatial reconstruction and narrative structures, help foster a more personal connection to history, especially important as the generation of eyewitnesses continues to pass away. Like these initiatives, the Santa Clara platform aims to combine spatial reconstruction with interactive storytelling to create a more engaging learning environment.

\section{Background}

Santa Clara 3D is the result of a close collaboration between the MEMORISE Project~\cite{memorise} and the Asociación Recuerdo y Dignidad~\cite{ard}. MEMORISE is a Horizon Europe-funded initiative that aims to preserve the memory of victims of Nazi persecutions and other forms of totalitarian violence through digital tools. As more victims of these persecutions pass away, developing new strategies to tell their stories becomes necessary to keep their memories alive. To achieve this, MEMORISE explores various digital technologies, aiming to make historical testimonies, documents, and sites more accessible and engaging to the public.
The Asociación Recuerdo y Dignidad is a volunteer-run association based in Soria, Spain, that is dedicated to recover historical memory of the Spanish Civil War and the Francoist repression. Their work includes archival research, locating and exhuming mass graves, and raising awareness of these histories from a human rights perspective. By providing historical and archival documents, testimonies and expert insights, the collaboration has been essential for the visual and narrative aspects of the reconstruction.

The Santa Clara 3D project focuses on the digital reconstruction and storytelling of the former concentration camp Santa Clara in Soria, Spain. The project demonstrates how lesser known or unavailable sites can be explored digitally, offering educational value and promoting awareness. During and after the Spanish Civil War, the Santa Clara complex served as a concentration camp. The site held mainly political prisoners under harsh conditions, suffering from overcrowding, lack of food and basic hygiene, forced labour and in some cases executions without a trial~\cite{rodriguez2011carceles}. Despite the site’s historical significance, it has received limited attention and remains relatively unknown to the public. This project seeks to contribute to the recognition of the site by digitally reconstructing the concentration camp and presenting its history through a web-based platform. The target audience includes educators, students, researchers, memory activists and others with an interest in history. The platform aims to make the site’s history accessible to a wide audience, encouraging exploration, reflection, and learning across generations.

\section{Methodology}

The development process reflects an approach inspired by participatory visualization design~\cite{janicke2020participatory}, which highlights the benefits of including stakeholders in the design process, by implementing feedback loops to ensure a satisfying final product. During the design process of Santa Clara 3D, data scientists of the MEMORISE team exchanged frequently with associates of the Asociación Recuerdo y Dignidad. Weekly feedback on prototypes has been essential to assure that the final product aligns with user needs.

\begin{figure}[t!]
 \begin{center}
 \includegraphics[width=0.6\linewidth]{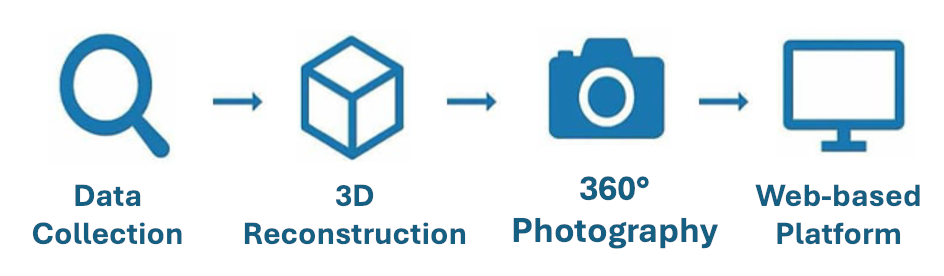}
 \caption{Methodological overview of the project steps, from collecting and transforming data to integrating it to the digital platform.}
 \label{fig:methodology}
 \end{center}
\end{figure}

\newpage
A combination of digital reconstruction, interactive media, and storytelling are key methods used to communicate the history of the Santa Clara Concentration Camp. The methodology integrates historical research, architectural reconstruction, and the development of a web-based platform to make a digital experience (see~\autoref{fig:methodology}).

\subsection{Data Collection}

To create a strong foundation and understanding of the project’s intentions, it was necessary to collect data to gain deeper insight into the history of Santa Clara. This included historical research not only about the site’s time as a concentration camp, but also its origins as a convent and its present-day condition. Since historical information about the site were limited, most data was gathered from historical articles and a range of archival materials, including old maps, architectural drawings, and testimonies. These documents were used to understand the function, use, and layout of the site over time.

In addition to remote archival research, a field trip to Soria was conducted, aiming to better understand the "domain situation" (in accordance to the nested model for visualization design by Munzner~\cite{munzner2009nested}). Our goal was to explore the city and visit historical sites that influenced events during the Spanish Civil War. By participating in guided tours around the city and inside the Santa Clara complex, valuable local insight into the site's historical evolution was gained. Furthermore, expert input was offered by a local archaeologist who has worked on excavations at the Santa Clara site. He provided detailed information about the layout, transformations and historical significance of the complex. Overall, the field trip made a significant contribution to both the digital reconstruction and the storytelling elements integrated into the platform.

\subsection{3D Reconstruction}

Based on the documentation gathered during the data collection phase, we developed a digital 3D reconstruction of the Santa Clara Concentration Camp. This reconstruction was based primarily on architectural plan drawings from the period when Santa Clara functioned as a concentration camp. The modelling was carried out using SketchUp~\cite{SketchUp}, a tool chosen for its accessibility and suitability for architectural modelling. A free version of SketchUp can be used for reconstruction purposes, however, novice users will face a steep learning curve in translating from historical architectural plans into accurate 3D representations. Santa Clara consisted of sixteen buildings during its time as a concentration camp, and each of these was modelled individually to create a comprehensive visualisation of the camp. The modelling process involved several key steps:
\begin{itemize}
 \item Loading the architectural plan drawings of each building into SketchUp
 \item Scaling each drawing to match the real-world dimensions
 \item Tracing the architectural outlines using vector lines to define walls and floor plans
 \item Adding architectural elements such as windows, doors and structural features
 \item Extruding the 2D plans into 3D models
 \item Applying wall and roof details based on photographic references
\end{itemize}

Both photographs taken on the site and historical ones from the period when the camp was active were used during the 3D reconstruction phase. The historical photographs were particularly useful for the modelling of stylistic features such as window and door types, with the aim of maintaining architectural consistency with the time.

\begin{figure}[ht]
 \begin{center}
 \includegraphics[width=\linewidth]{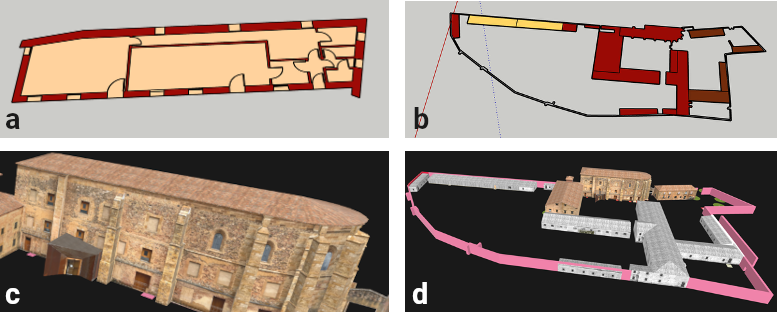}
 \caption{3D reconstruction of Santa Clara: (a) 2D floor plan of a single building from the Santa Clara concentration camp, created in SketchUp, (b) 2D overview of the entire camp layout, showing the placement and dimensions of all 16 buildings, (c) 3D model of one reconstructed building, including architectural features and texturing based on historical and photographic references, and (d) 3D overview of the entire camp reconstruction. The image highlights the visual contrast between buildings with and without applied wall textures.}
 \label{fig:reconstruction}
 \end{center}
\end{figure}

\newpage
To illustrate parts of the modelling process,~\autoref{fig:reconstruction} presents both 2D and 3D representations of the reconstruction in SketchUp.
\autoref{fig:reconstruction}a shows a 2D floor plan of a single building, while~\autoref{fig:reconstruction}b illustrates an overview of the entire camp layout. Correspondingly, ~\autoref{fig:reconstruction}c and d display 3D models of one of the buildings and the complete camp, respectively. In the 3D models, the application of wall and roof textures is visible, with~\autoref{fig:reconstruction}d particularly highlighting the contrast between textured and non-textured surfaces. All modelling decisions were guided by the aim of achieving both spatial and historical accuracy, to make the reconstruction as authentic as possible, based on the available sources.  

\subsection{360° Photography}

In the process of developing the digital twin of the present state of Santa Clara, different methods were considered, including photogrammetry~\cite{schenk2005introduction} and LiDAR scanning~\cite{raj2020survey}. However, since only three buildings remain, none of which contain original architectural elements from the concentration camp period, a detailed scan of the current structures was deemed less relevant for a project focused on historical reconstruction.
As an alternative, the site's present-day state was documented using low-cost 360° photography~\cite{matzen2017low}. This method is quick, easily accessible, and provides sufficient spatial awareness while creating a visual and navigable experience.

During the field trip to Soria, the 360° photographs were captured on site. Polycam~\cite{polycam} was chosen to create the images, as it provides an easy-to-use platform with consistent results. The photographs were taken using an iPad Pro mounted on a tripod to ensure stable movement, making the images more consistent and clearer. Photographs were captured throughout the entire complex, allowing users to explore the site virtually by navigating between images.
The 360° photographs were then processed and integrated into the platform using a three.js viewer~\cite{dirksen2015three}. This allows users to click and look around freely, providing a realistic impression of how the site appears today. Including these images helps establish a stronger connection between the present-day landscape and the historical events that took place, supporting users in understanding the contrast between past and present.

\subsection{Development of Web-based Platform}

To convey the history of Santa Clara and enable public exploration, a web-based platform was created. It was developed using HTML, CSS, and JavaScript, and incorporates libraries such as MapLibre~\cite{MapLibre} for the interactive map and Three.js~\cite{dirksen2015three} for rendering the 3D models.
The platform was designed as an immersive way of exploring the history of the Santa Clara site through different technical solutions. By enabling non-linear engagement with Santa Clara’s layered history, the project takes inspiration from Mitchell Whitelaw’s concept of the Generous Interface~\cite{whitelaw2015generous}. 
This concept promotes rich, browsable overviews rather than limiting access through search functions or hierarchical menus. 
Through the integration of an interactive map containing markers with distinguishable icons for different media types (3D model, 360° photographs, video, historical photographs), the platform encourages curiosity and invites exploration.

A timeline slider allows users to switch between three historical layers of the site: (1) its origins as a convent, (2) its use as a concentration camp, and (3) its present-day condition. This multi-layered approach, combined with spatial navigation, helps users develop a deeper understanding of the site's transformation over time. The design avoids predetermined storytelling paths and instead allows users to construct their own narrative journeys through movement, image exploration, and interaction with historical content.

To simplify updates and future changes, the interactive markers were implemented using separate GeoJSON files. These specify where each marker is located and what it contains. Each marker is categorized by their content - such as “3D models,” “360° view,” “Historical info,” or “Video” and represented by custom icons. When a marker is selected, a pop-up window appears displaying a media and a descriptive text. This approach is made to ensure that each piece of content can stand independently, while still contributing to a broader historical narrative.
The integration of the 3D reconstruction and the 360° photographs was achieved by exporting the SketchUp models and the Polycam photos into GLB-format and rendering them through use of Three.js library. This allows users to pan, zoom, and explore the virtual content in detail. The interface was developed to offer a meaningful way for users to engage with the memory and legacy of the camp through interaction, spatial awareness, and layered historical context.

\begin{figure}[ht]
 \begin{center}
 \includegraphics[width=.9\linewidth]{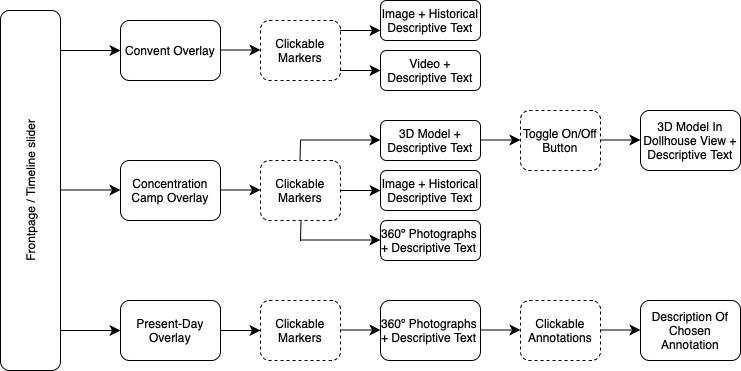}
 \caption{Flowchart illustration of possible user interactions on the platform.}
 \label{fig:interactions}
 \end{center}
\end{figure}

\section{Results}

Santa Clara 3D is an interactive web-based platform that allows users to explore the history and the spatial layout of the former Santa Clara Concentration Camp. By integrating 3D models, 360° photographs, and historical information, the platform creates a layered, explorative digital experience. In the following, the interface is presented through screenshots, and a flowchart illustrates possible user interactions (see~\autoref{fig:interactions}).

\begin{figure}[ht]
 \begin{center}
 \includegraphics[width=\linewidth]{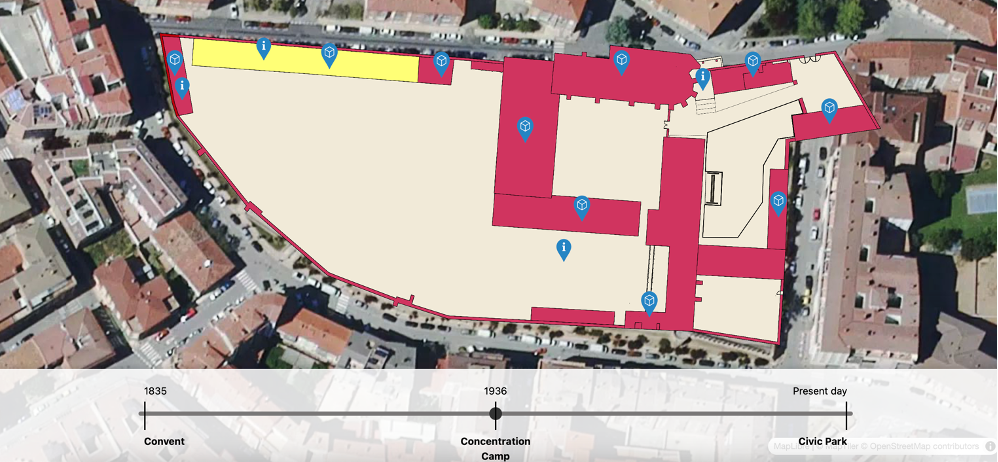}
 \caption{\textit{Concentration Camp Layer (1936)} – A reconstructed site plan overlays the map, showing the Santa Clara complex during its use as a concentration camp. Markers open 3D models and historical descriptions of the buildings, enabling users to explore the camp's structure and use.}
 \label{fig:camp_layer}
 \end{center}
\end{figure}

\begin{figure}[h]
 \begin{center}
 \includegraphics[width=\linewidth]{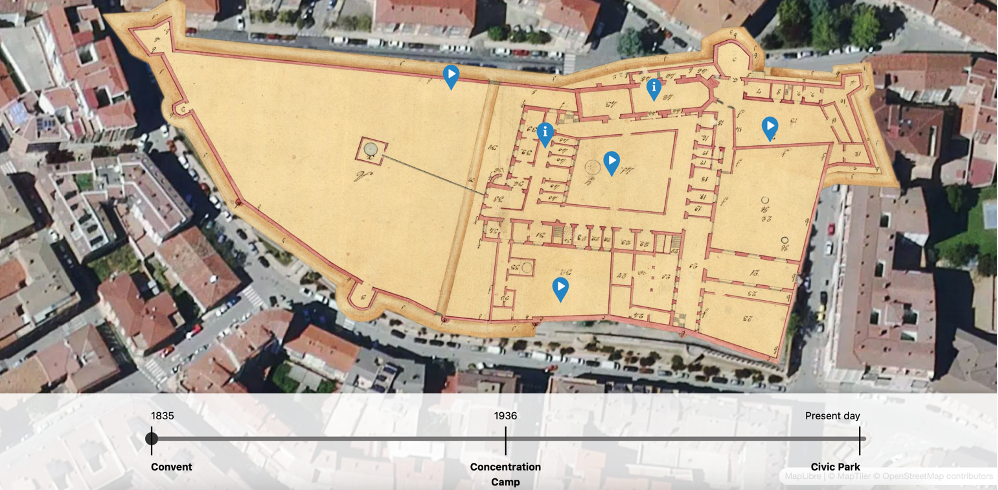}
 \caption{\textit{Convent Layer (1835)} – A hand-drawn historical map depicts the original layout of the Santa Clara convent. Informational and video markers provide insight into the site's religious function and early architecture.}
 \label{fig:convent_layer}
 \end{center}
\end{figure}

Upon loading the platform, users see an interactive map centred on the Santa Clara complex. The map allows zooming and panning to navigate the area and includes a timeline slider at the bottom of the interface. This slider allows users to toggle between three historical layers: the site's origin as a convent, its use as a concentration camp, and its present-day condition. In the concentration camp layer, a historical site plan from that period is applied as a map overlay (see~\autoref{fig:camp_layer}), and a hand-drawn map from 1835 serves as the base layer in the convent layer (see~\autoref{fig:convent_layer}). A smooth fade transition between layers helps users visually observe the architectural changes over time.

Each layer contains its own set of markers with custom icons indicating different types of media: 3D models, 360° photographs, historical information, and video. When clicking a marker, a pop-up window appears, displaying the media on the left and a descriptive text on the right. Navigation arrows allow users to move between markers, and an exit button returns them to the main map.


\subsection{360° Photograph Annotations}
\begin{figure}[ht]
 \begin{center}
 \includegraphics[width=\linewidth]{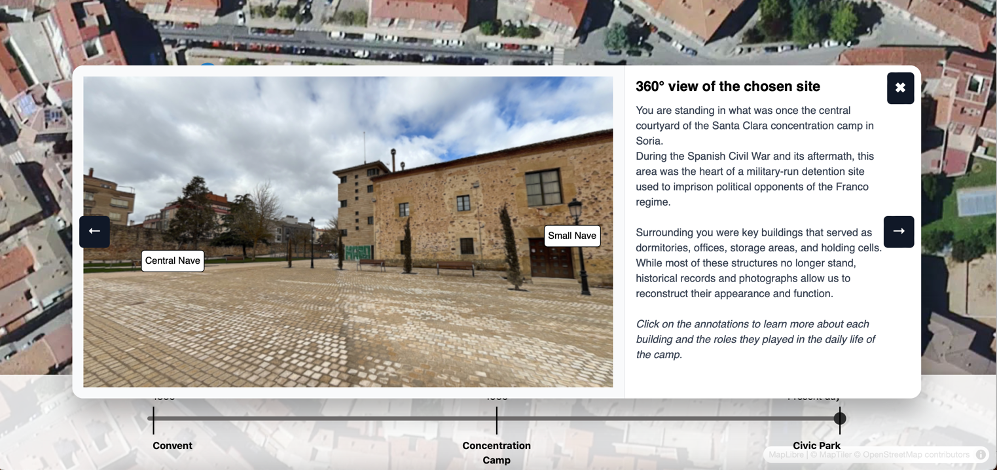}
 \caption{\textit{360° Photograph with Annotations} – A panoramic image of the Santa Clara site showing the existing Small Nave and the former Central Nave, with clickable annotations that provide historical descriptions of each structure.}
 \label{fig:360_annotations}
 \end{center}
\end{figure}

The present-day and the concentration camp layer includes 360° photographs taken on-site using Polycam. These images allow users to virtually “stand” inside the complex and explore the surroundings in a panoramic view. Annotations within the 360° images mark the locations of current buildings and buildings that once existed during the concentration camp period. Clicking an annotation reveals a brief description of that building’s former function (see~\autoref{fig:360_annotations} for an example showing the Small Nave and the former Central Nave).

\subsection{3D Model Annotations}
\begin{figure}[ht]
 \begin{center}
 \includegraphics[width=\linewidth]{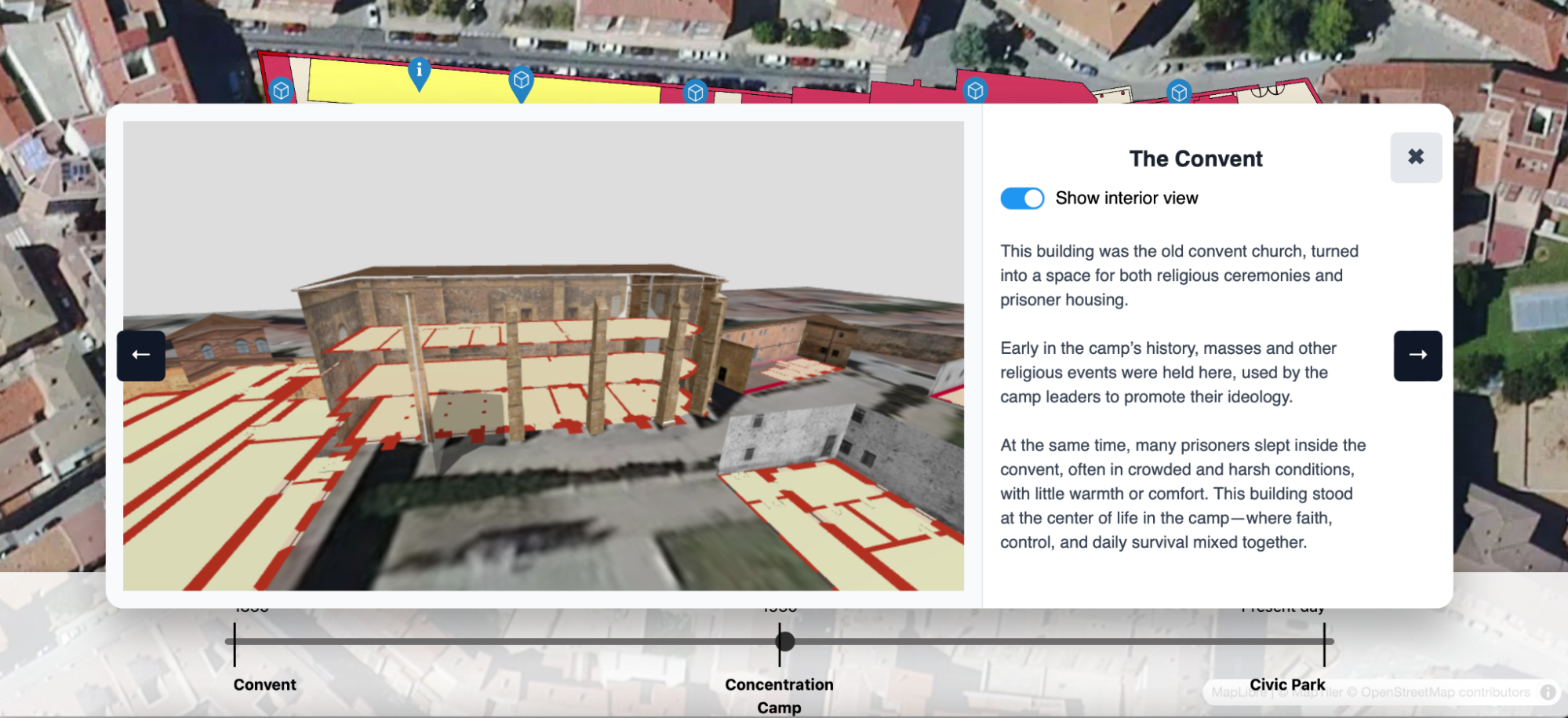}
 \caption{Dollhouse-view of the 3D reconstruction of the Convent.}
 \label{fig:3D_annotations_dollhouse}
 \end{center}
\end{figure}

Markers labelled with a 3D icon are located on several buildings in the concentration camp overlay and indicate buildings that have been digitally reconstructed. When selected, the pop-up displays a 3D model of the Santa Clara complex with the chosen building in focus, along with a descriptive text. A toggle button reveals a dollhouse-style view of the building, allowing users to examine the interior layout (see~\autoref{fig:3D_annotations_dollhouse}).

\begin{figure}[ht]
 \begin{center}
 \includegraphics[width=\linewidth]{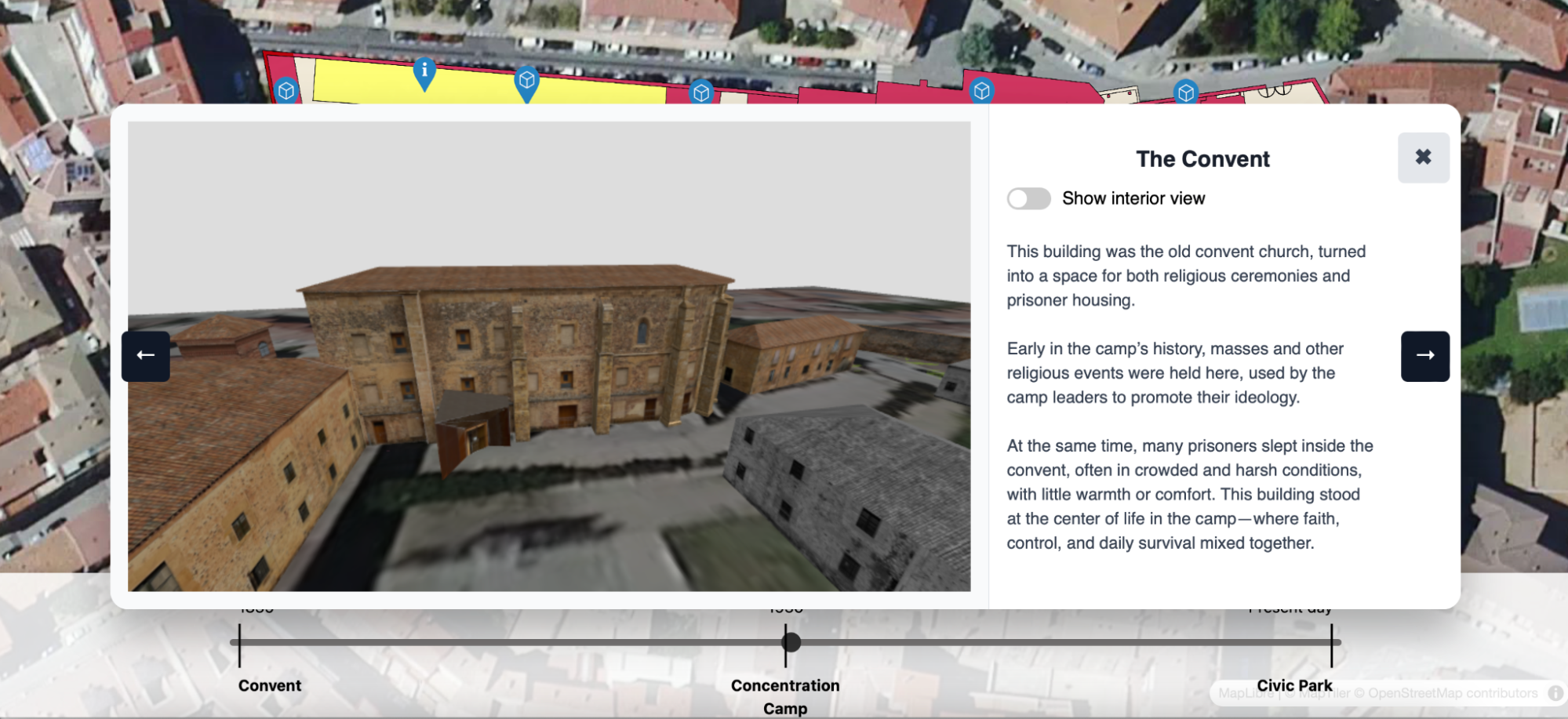}
 \caption{3D reconstruction of the Convent.}
 \label{fig:3D_annotations_convent}
 \end{center}
\end{figure}

To distinguish between buildings that no longer exist and those that still stand today, two different texture styles are applied: photographic textures based on current site visits and historical photos for existing structures, and monochrome textures for non-existing buildings. The models are fully interactive, allowing users to rotate and zoom for better understanding of the site’s spatial layout.

\subsection{Video Annotations}
Markers featuring a “play” icon open embedded videos and a descriptive text when clicked. These short clips, which must be manually started, show a local archaeologist discussing the site’s historical significance and providing context primarily about its use prior to becoming a concentration camp.


\subsection{Photograph Annotations}
\begin{figure}[ht]
 \begin{center}
 \includegraphics[width=\linewidth]{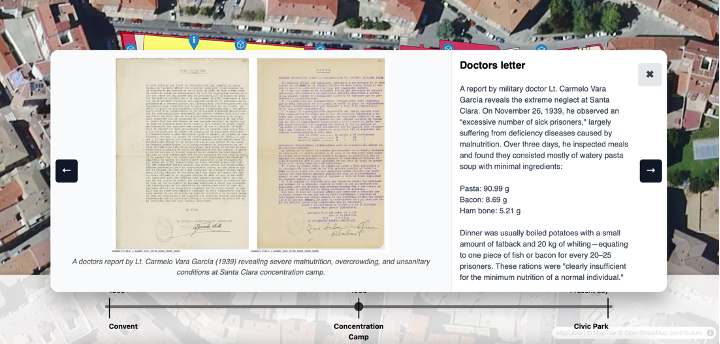}
 \caption{Pop-up of the concentration camp layer showing a 1939 medical testimony by a doctor who evaluated camp conditions, accompanied by contextual information.}
 \label{fig:photo_annotations_doctor}
 \end{center}
\end{figure}

Markers with an “i” icon appear in both the convent and concentration camp layers. These provide archival photographs and historical context on key topics such as “Living Conditions”, “Types of Prisoners”, and “Forced Labour”. 
\autoref{fig:photo_annotations_doctor} illustrates a marker presenting a scanned letter from a doctor who evaluated the camp’s conditions in 1939. Additionally, the “Forced Labour” marker includes a zoom-out feature that reveals two off-site locations associated with prisoner labour.


\section{Discussion}

In the development of the platform, many design choices were considered to accommodate the implementation of different media types, interactive tools, and historical information.

\newpage
\subsection{Iterative Design Process}

\begin{figure}[ht]
 \begin{center}
 \includegraphics[width=\linewidth]{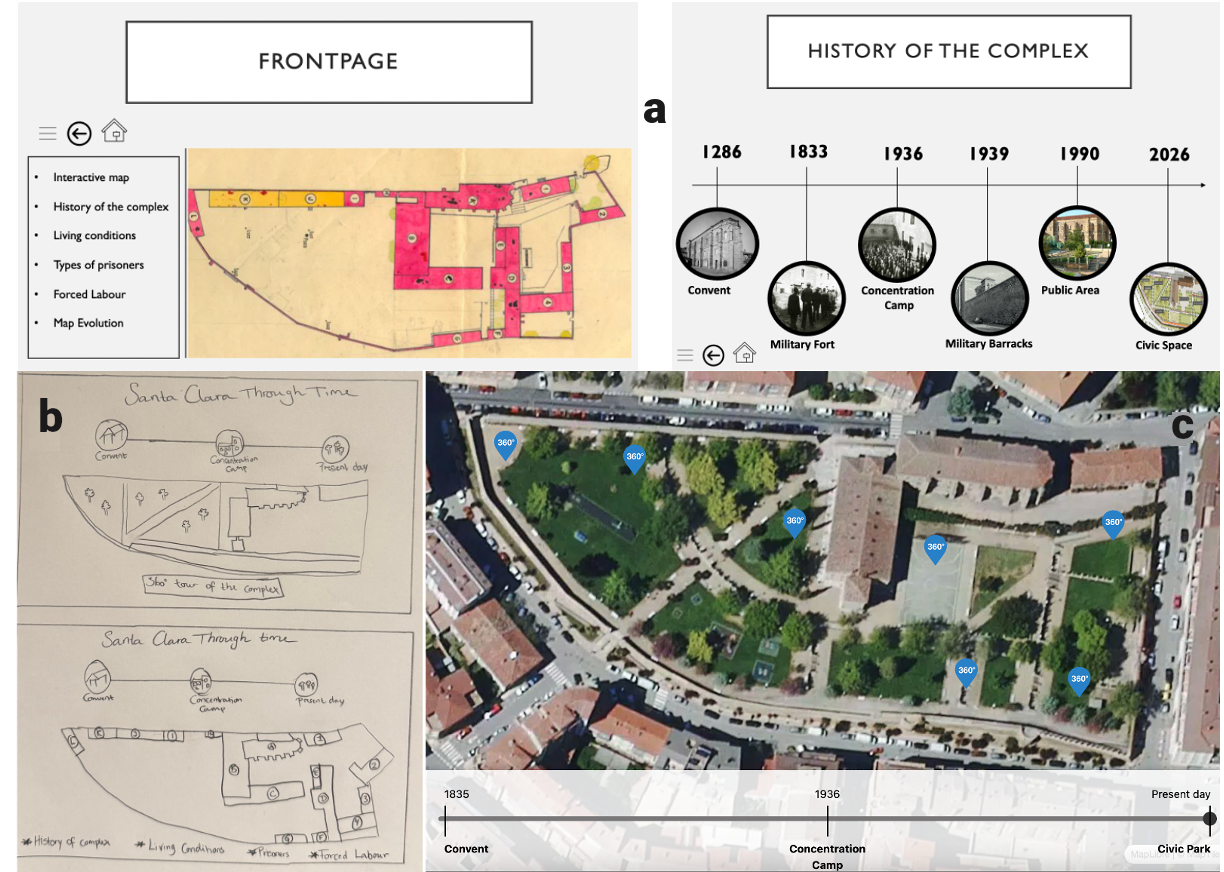}
 \caption{Sketches from the design process showing the evolution of the platform interface. (a) Early concept with a central 3D model and separate menu. (b) Sketch combining the timeline slider with the main map view, while still including a menu for thematic content. (c) Final design featuring an interactive satellite map with bottom timeline and interactive markers for open-ended exploration.}
 \label{fig:sketches}
 \end{center}
\end{figure}

As part of the design process, the interface evolved through several sketches (see \autoref{fig:sketches}), aiming to improve user interaction and exploration of the site. In early drafts, a front page focused on the 3D model of Santa Clara with a separate menu for exploring different topics, like the timeline, was imagined (see \autoref{fig:sketches}a). However, this structure lacked the ability to show the historical depth of the site, and the layout felt too separate.

In later sketches (see \autoref{fig:sketches}b), the design shifted toward combining all components into a single main view. The idea of placing the timeline slider directly on the map was implemented, allowing users to move through different historical periods of the site. The platform still included a menu that offered topics such as "Living Conditions" and "Forced Labour". The intention of keeping everything in one view was to make the experience more intuitive and support user-led exploration.

In the final design (see \autoref{fig:sketches}c), the timeline is placed at the bottom of a satellite map, and the menu was replaced by clickable markers that offer an open-ended way of interacting with the site's content. Overall, the design process focused on finding a balance between clarity, engagement, and historical reflection. Each stage helped shape a platform that emphasizes the importance of showing change over time and keeping the memory of these places alive.      

\subsection{Historical Authenticity}
The intention of using digital tools to contribute meaningfully to the remembrance of historical sites associated with difficult pasts, such as the Spanish Civil War and Francoist repression, was a central motivation throughout the creation of this platform. The project aimed to demonstrate how digital technologies can help communicate such complex topics while also encouraging exploration, reflection, and emotional engagement.

A key design strategy was the use of a timeline slider, which allows users to toggle between different historical phases of the Santa Clara complex. Beyond simply visualising change, this feature highlights the site's transformation over time and serves as a narrative device. By allowing users to shift between temporal states, the platform illustrates how historical sites are overwritten and transformed. This intends to invite users to reflect on how histories can be hidden or forgotten, underscoring the importance of remembrance.
The monochrome textures applied to the reconstructed 3D buildings reinforce this theme of absence. The use of grayscale signals that these buildings belong to the past and emphasizes that their visualisation is based on interpretation and educated guesses, rather than certainties. The contrast between grayscale and full-colour textures visually conveys the idea of memory and loss, aligning with the idea of the history being partly lost or hidden.

During the development, custom-made markers were implemented to distinguish between different types of media. Options such as numbering the markers to suggest a fixed story order, or using color-coded categories, were considered. However, the design was chosen to allow the users the ability to explore the content in their own way. The open-ended design intends to encourage personal reflection rather than guiding users through a fixed narrative~\cite{whitelaw2015generous}.  

One central challenge was the limited availability of reliable information and resources about Santa Clara. Much of the history from this period remains undocumented, or was actively erased, which made it difficult to ensure historical accuracy. As a result, some compromises had to be made, both in the historical content and the design of the platform. This was especially true in the 3D reconstructions, where certain visual details, like building textures or architectural features, had to be based on educated assumptions or comparisons with other structures known to have existed in the camp. These choices inevitably introduce some degree of uncertainty.

Being developed in close collaboration with the MEMORISE project, it was important that the platform aligned with the organization’s goals of promoting digital memory activism. Although this platform cannot replace a physical memorial or a historical archive, it offers users the opportunity to explore, interact, and reflect at their own pace. It is the hope that this will provide an intriguing experience especially for younger generations or people unfamiliar with the history. 
In the end, this platform aims not only to present the history of Santa Clara, but also to serve as an example of how digital technologies can contribute to memory work, public history, and historical awareness.

\subsection{Evaluation}

The Asociación Recuerdo y Dignidad played a crucial role in strengthening the platform’s historical foundation by supplying archival material and validating historical narratives. Their input on content related to forced labour, types of prisoners, and living conditions helped ensure the platform presents a respectful and accurate portrayal of the site's past.

As part of the final development stage, a meeting was held with three representatives from the Asociación Recuerdo y Dignidad to present and evaluate the final prototype. This expert evaluation confirmed the project’s clarity and value, while also proposing improvements such as the addition of an information box on each overlay to introduce and clarify the time period shown.

To further improve the platform, a more comprehensive evaluation should ideally be conducted. By presenting the platform in a controlled setting, such as a physical exhibition, researchers could collect data through direct observations, user interviews, and logging of user behavior across devices. This approach would provide valuable insights into usability and emotional engagement, ultimately enhancing the platform’s effectiveness as a tool for historical reflection and education.

This collaborative process ensured that the final platform met its stated objectives, offering a meaningful, research-based, and interactive experience that contributes to the ongoing remembrance of this historical site.

\subsection{Limitations}

While the platform successfully meets many of the intended goals, several limitations have influenced the outcome of the project. Working under time constraints inevitably required certain compromises, limiting the final product.
The available archival materials, testimonies, and visual references related to the Santa Clara Concentration Camp are very limited. Much of the historical information was retrieved from local articles, and many assumptions about the purpose of specific buildings were based on educated guesses - grounded in historical floor plans and typical uses of similar spaces. Because of the limited material, certain narrative and reconstruction choices relied on interpretation rather than definitive evidence. Although this was acknowledged and addressed throughout the project, it affected the amount of detail that could be included and reduced the level of historical precision in some areas. Additionally, all of the historical documents provided by Asociación Recuerdo y Dignidad were in Spanish, requiring the use of online translation tools to interpret the content. While these translations were handled carefully, they may have introduced slight inaccuracies or misunderstandings that affected the level of detail in some areas.

The collaboration with Asociación Recuerdo y Dignidad was crucial for providing historical context. However, as this organization operates primarily through voluntary work, communication sometimes lagged due to limited availability or resources. This occasionally led to improvised solutions, which may have affected the depth or accuracy of some aspects of the project.

\subsection{Future Work}

This project contributes to a great first step in using digital tools to engage with the history of Santa Clara Concentration Camp. However, further developments could improve Santa Clara 3D.

\begin{itemize}
 \item \textbf{User testing:} Throughout the development of the platform, weekly feedback has been essentiel in shaping the final design choices. However, to reach a broader and more diverse audience, structured user testing could be considered. By observing how different users interact with the platform, improvements in general areas like navigation and overall user-friendliness could be identified. This feedback could lead to the development of a more tailored user experience, enhancing the platform for different user groups.
 \item \textbf{Tutorial:} The implementation of a short introduction or tutorial on how to use the platform could help users understand the platform’s structure and tools more easily. This would especially be useful for user groups unfamiliar with digital heritage platforms or 3D interfaces.
 \item \textbf{Expanding content:} Integrating more archival materials, oral histories, personal testimonies or other resources from families of victims or local residents would greatly enhance the overall user experience. These additions would reinforce the memory of the narratives and add more content to the historical experience.
 \item \textbf{Performance:} Future work could focus on improving the platform’s overall technical performance. This might include implementing lower-resolution 3D models to reduce loading times or optimising the code for better cross-device compatibility—especially for mobile devices. Although the platform is currently accessible on smartphones, mobile-specific design adjustments are needed to create a fully usable mobile experience.
 \item \textbf{Extended Reality:} Future development could include incorporating virtual reality (VR) to allow users to explore the reconstructed site in a more immersive and intuitive way. Augmented reality (AR) could also be relevant, enabling users to visualise former buildings digitally while physically present on the site. These technologies could create deeper emotional significance and further strengthen the educational potential of the project.
 \item \textbf{Multi-language support:} As this project was developed in collaboration with a Spanish organisation, it would make sense to provide a Spanish translation of the platform. Making the platform multilingual could help attract a more diverse audience and broaden accessibility.
\end{itemize}

Overall, these suggestions for future improvements could enhance the platform’s accessibility, engagement, and educational impact, resulting in a deeper and more immersive exploration of the history of Santa Clara.

\section{Conclusion}

The main goal of this work was to explore how the implementation of digital tools could enhance the remembrance and interpretation of neglected historical sites. It focused on Santa Clara Park in Soria, Spain, which functioned as a concentration camp during the Spanish Civil War and now shows only minimal traces of its harsh past. Through 3D reconstruction, 360° photography, and historical research, a web-based platform was created to highlight the site's historical significance.
The platform demonstrates how digital tools can offer new ways to engage with difficult historical topics. By implementing features such as a timeline slider and layered map overlays, users are invited to explore the site's temporal evolution and reflect on what has been forgotten over time. Design choices, such as the use of custom markers and monochrome textures for non-extant buildings, emphasize historical transformation, spatial change, and emotional engagement.
While limitations affected the level of detail and historical precision, the platform offers a strong foundation for future development. Suggestions for further improvement include structured user testing, content expansion, technical optimisation, and the possible integration of VR or AR technologies.

The project highlights the value of digital heritage methods as meaningful tools for memory work. By combining digital reconstruction with an open-ended, exploratory interface and historical narratives, the platform invites a broader reflection on how digital tools can shape the way we remember the past. One of the main contributions of our work is the development of a low-cost pipeline that can easily adapted to reconstruction and storytelling projects for other historical sites.

%
%
%
%
\newpage
\bibliographystyle{splncs04}
\bibliography{references.bib}

\end{document}